\documentclass[draftcls,onecolumn,oneside,12pt]{IEEEtran}
\usepackage{times,subfigure}
\usepackage{graphicx,algorithm,algorithmic,multirow,flushend}
\usepackage{pgf,tikz}
\usepackage{graphicx,multirow}
\usepackage{latexsym,color,float,amsmath,amssymb,amsthm}
\IEEEoverridecommandlockouts

\theoremstyle{definition}

\theoremstyle{plain}

\theoremstyle{remark}

\newcounter{tmp-counter}

\begin{document}

\title{Link Scheduling for Multiple Multicast Sessions in Distributed Wireless Networks\thanks{Manuscript received December 16, 2012; revised January 25 and February 25, 2013; accepted March 20, 2013. The associate editor coordinating the review of this letter and approving it for publication was C. Cordeiro.}\thanks{The author is with the Department of Computer and Communication Engineering, University of Thessaly, Volos, 38221, Greece (anargyr@ieee.org).}\thanks{The author would like to acknowledge the support from the European Commission through the Marie Curie Intra-European Fellowship WINIE-273041 and the STREP project CONECT (FP7ICT257616).}}%

\author{\IEEEauthorblockN{Antonios Argyriou,~\IEEEmembership{Member,~IEEE}}}


\maketitle%


\markboth{IEEE Wireless Communications Letters, \today}{Link Scheduling for Multiple Multicast Sessions in Distributed Wireless Networks}

\begin{abstract}
In this letter we investigate link scheduling algorithms for throughput maximization in multicast wireless networks. According to our system model, each source node transmits to a multicast group that resides one hop away. We adopt the physical interference model to reflect the aggregate signal to interference and noise ratio (SINR) at each node of the multicast group. We present an ILP formulation of the aforementioned problem. The basic feature of the problem formulation is that it decomposes the single multicast session into the corresponding point-to-point links. The rationale is that a solution algorithm has more flexibility regarding the scheduling options for individual nodes. The extended MILP problem that also considers power control is solved with LP relaxation. Performance results for both the ILP and MILP problems are obtained for different traffic loads and different number of nodes per multicast group.
\end{abstract}


\section{Introduction}
\label{section:introduction}
\begin{figure}[t]
 \begin{center}
  \includegraphics[scale=0.345]{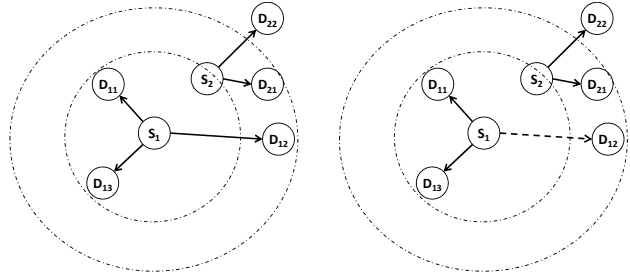}\\
 \end{center}
\caption{Example network with two concurrent multicast sessions. When both nodes multicast during the same slot, then they both interfere to at least one of the destinations of the other multicast group (left). When the transmission power of $S_1$ is reduced and it can only transmit reliably to $D_{11}$,$D_{13}$ then the result is that $S_2$ can multicast reliably to both its destinations (right). The optimal activations of the links that compose each multicast session is addressed in this work.}
\label{fig:multicast-sched-problem}
\end{figure}

Wireless multicasting poses significant technical challenges that have attracted considerable amount of research work for several years. The problem is more relevant than ever due to the widespread use of WiFi-enabled mobile devices that need to transmit high volumes of data to several users. Smart-phones with multicasting capabilities are envisioned as one of the key adopters of such technologies especially for the distribution of high-quality locally captured video data (at the multicast source). To this aim there are notable practical efforts for wireless multicasting in WiFi networks~\cite{chandra09,dujovne06}. In this letter we are concerned with an environment where multicasting is routinely adopted for wireless transmission among nodes in a distributed network (see Fig.~\ref{fig:multicast-sched-problem}). This means that several nodes may need to concurrently use multicast communication.

In wireless multicast one of the key problems is the impact of heterogeneous destinations on the performance of the entire multicast session~\cite{dujovne06}. A destination that belongs to the multicast group, but is characterized by the worst channel quality among all the destinations, will be the bottleneck of the multicast communication since it will require increased transmit power and retransmissions. A scenario beyond a single multicast session needs to consider a wireless network where several muticast sessions are active in the same space, time period, or even frequency band. Therefore, it is also possible that a multicast session generates interference to the destinations of the rest of the multicast sessions. This creates an additional problem for the performance of multicast in the wireless network. Fig.~\ref{fig:multicast-sched-problem} depicts this situation where sources $S_1,S_2$ that multicast to their respective groups $D_{11},D_{12},D_{13}$ and $D_{21},D_{22}$ respectively. In this case $S_1$ must increase the transmit power to a level that the packet is decodable also from $D_{12}$. However, this decision will generate interference to node $D_{21}$ and it may render undecodable the second multicast transmission. The same is true for decisions made by $S_2$ (transmission range is not shown).

The problem we address in the context of the previously described scenario is the following. Given a network with a set of next-hop multicast sessions how can we schedule the wireless multicast transmissions so that throughput is maximized? Significant performance improvements are observed and originate from a re-formulation of the integer linear problem (ILP) that describes the throughput maximization problem for this scenario. Subsequently, the ILP problem formulation is extended to a mixed ILP (MILP) that also considers power control and is solved with LP relaxation and a heuristic that exploits the multicast nature of the problem. The only disadvantage of the proposed approach is a higher number of optimization variables.

Scheduling multicast transmissions in wireless networks has attracted a certain amount of work even a few years earlier. Wieselthier~\emph{et. al} had shown in~\cite{wieselthier00} the first results that indicated that with proper power allocation (called the broadcast incremental power algorithm) it is possible to minimize interference for multicast sessions. Heuristic joint power control and scheduling algorithms were also studied by Wang~\emph{et. al} in~\cite{wang03}. This work is of particular interest since the authors focused on scheduling for wireless multicasting but with the objective to minimize power consumption. For a given group of multicast sources and their corresponding destinations, the power optimization problem was formulated as a MILP. The optimal values of the transmit power were evaluated so that the SINR requirements at the receivers were fulfilled while the total power expenditure was minimized. In more recent works Gopinathan~\emph{et. al} presented a model for optimal multicast in multi-channel multi-radio wireless networks under the assumption that the channel assignment is static~\cite{gopinathan09}. Krishnan~\emph{et. al} focused on the problem of identifying the optimal multicast trees by considering also the next hops in multi-hop wireless ad hoc networks~\cite{krishnan10}. More theoretical studies driven from an information-theory perspective, focus on exploiting the broadcast advantage for multiple sessions~\cite{cui10}. In~\cite{cui10} multicasting is considered in a system that employs network coding and allows cross-layer interactions while the authors follow a utility optimization scheme for calculating the optimal source rate.

\section{System Model and Initial Problem Formulation}
\label{section:system-model}
We study a network model where a set $\mathcal{S} = \{S_1,S_2,...,S_N\}$ of sources want to communicate with a number of multicast destination nodes that are denoted as the set $\mathcal{D}$. Each source multicasts to $D$ nodes. The definition of a multicast link in this letter is extended and it involves a number of nodes that is equal to the size of the multicast group. We refer explicitly to  a link between two nodes as a point-to-point link. The complete network is modeled with a directed graph $F(\mathcal{U},\mathcal{V})$, where $\mathcal{U}$ and $\mathcal{V}$ are the set of point-to-point directional links, and the set of nodes, respectively ($\mathcal{V} = \{\mathcal{S},\mathcal{D}\}$). We also use the conflict graph of $F$ that is defined as $G(\mathcal{U},\mathcal{V})$, and contains the interfering relationships among the $N\times D$ point-to-point links in the network. Each vertex in the conflict graph represents a wireless link in the network, and there is an edge between two vertices if and only if the links represented by the vertices conflict (i.e. they interfere with each other and simultaneous transmission is impossible). On the other hand, a clique in the conflict graph represents a group of links that cannot transmit concurrently, and hence they must access the channel exclusively.

\subsection{Initial MILP Formulation}
In the most closely related work to this letter the objective was power minimization for a wireless multicast scenario~\cite{wang03}. We follow a similar approach for defining the initial multicast optimization problem but in our case we consider throughput maximization. In this formulation, the binary optimization variable $x^{t}_{i}$ indicates whether a transmission from the source of the multicast group $i$ occurs in slot $t$. $T$ is the maximum number of slots, $P^t_i$ is an optimization variable that corresponds to the transmit power of source $i$ during slot $t$, $\beta$ is the SINR decoding threshold at the destination, $\sigma^2$ is the AWGN variance, $\gamma_{ij}=1/{d^a_{ij}}$ where $d_{ij}$ is the distance between source $i$ and destination $j$ while $a$ is usually set in a value between 3 and 4. The actual problem formulation named $\mathbf{MC-ALL}$ is given below:
\begin{eqnarray*}\label{eqn:problem5}
&& \max_{x^{t}_{i}, P^{t}_i} \frac{1}{T} \sum_{t=1}^T  \sum_{i=1}^{N} D_{i}x^{t}_{i} \\
&& \frac{P^t_{i}\gamma_{ij}+(1-x^{t}_{i})\Delta}{\sigma^2+\sum_{k\in\mathcal{S}-\{i\}} P^t_{k}\gamma_{kj}} \geq \beta,  \forall j\in \mathcal{D}_i,  i\in \mathcal{S},  t\in T~(1)\\
&& \sum^{T}_{t=1} x^{t}_{i} \geq 1,\forall i\in \mathcal{S}~(2),\quad \sum^{T}_{t=1} P^{t}_{i} \leq P^{max}_{i},\forall i\in \mathcal{S}~(3), \\
&& 0\leq P^{t}_i \leq P^{t,max}_{i}x^{t}_{i}~(4),\quad x^{t}_{i} \in \{0,1\}~(5)
\end{eqnarray*}
The objective is to maximize the throughput by increasing the number of scheduled sources and after taking into account as a weight the number of nodes in the multicast group of each source $D_i$. Next, constraint (1) is the SINR constraint. In (1) $\Delta$ is a large number needed for ensuring that the SINR constraint is satisfied when the respective link is not scheduled. In our performance evaluation it is set to a value higher than the maximum possible SNR in the network. The key observation from the formulation in (1) is that it ensures that the SINR constraints are satisfied for all the multicast destinations of each source. The remaining constraints (2),(3),(4) ensure first that each source is scheduled at least once during the $T$ slots, second that the transmitter power is limited according to a maximum value $P^{max}_{i}$ for the complete set of $T$ slots, and third that that the transmitter power $P^{t,max}_{i}$ for a specific slot is limited depending on the transceiver specifications. A note should be made here that will set the stage for our proposed optimization approach. Whenever the optimal solution cannot be found for the given set of multicast sources, strong interferers are completely eliminated (and this means complete multicast groups cannot be scheduled in that slot, i.e. $x^{t}_{i}=0$). This is a key disadvantage of the $\mathbf{MC-ALL}$ approach.

\section{Throughput Maximization in Multicast Wireless Networks}

\subsection{Proposed MILP Formulation}
The last observation we made for the behavior of $\mathbf{MC-ALL}$ stems from one critical detail. In the previous formulation a specific multicast group is treated as a single schedulable link/entity. Avoiding this limitation is necessary for allowing more flexible scheduling decisions. To this aim we relax the requirement that with a single broadcast transmission from a source, all the associated multicast destinations must receive the packet. We introduce now the optimization variable $x^t_{ij}$ that indicates the activation of a single point-to-point link from multicast source $i$ to a specific destination $j$. The optimization variables for the transmitter power remain the same and are denoted again as $P^t_{i}$. The $\mathbf{DMC-OPT}$~(Dynamic Multicast) problem allows the source \emph{to activate dynamically a subset of its point-to-point links that compose a multicast link in a specific slot $t$ and not the complete set in order to achieve throughput optimality}. For the remaining non-scheduled destinations we introduce constraints that ensure that they are scheduled during a number of slots. The detailed formulation is named $\mathbf{DMC}-\mathbf{OPT}$ and is given below:
\begin{eqnarray*}\label{eqn:problem5}
&& \max_{x^{t}_{ij}, P^{t}_i} {\frac{1}{T} \sum_{t=1}^T  \sum_{i=1}^{N}  \sum_{j=1}^{D_i} x^{t}_{ij}}\\
&& \frac{P^t_{i}\gamma_{ij}+(1-x^{t}_{ij})\Delta}{\sigma^2+\sum_{k\in\mathcal{S}-\{i\}} P^t_{k}\gamma_{kj}} \geq \beta, \forall j\in \mathcal{D}_i, i\in \mathcal{S},  t\in T~(1)\\
&& \sum^{D_i}_{j=1} x^{t}_{ij} \leq D_i, \forall i\in \mathcal{S}~(2), \quad x^{t}_{ij} \in \{ 0,1 \}~(3) \\
&& \sum^{T}_{t=1} x^{t}_{ij} \geq B_{i,j}, \forall j\in D_i, i\in \mathcal{S}~(4)\\
&& \sum^{T}_{t=1} P^{t}_{i} \leq P^{max}_{i}~(5),\quad P^{t,min}_i \leq P^{t}_i \leq P^{t,max}_{i}x^{t}_{ij}~(6)\\
\end{eqnarray*}
Constraint (1) ensures that each of the $D_i$ destinations, that are counted with the subscript $j$ and correspond to the multicast group originating from source $i$, must have the SINR higher than the required threshold $\beta$. The next constraint (2) essentially says that when a node multicasts, this action corresponds to the activation of a number of source-destination point-to-point links that their number must be less or equal to the number of multicast destinations $D_i$. With this constraint it is possible that not all point-to-point links are \textit{activated}. Although a signal from a transmission might be received at every network node, with the term activation we mean the selection of a proper power level so that the corresponding destination can decode the packet (i.e. the SINR is above $\beta$). A valid 0 or 1 value for the activation of a particular link is ensured with constraint (3). It is important to clarify constraint (4) that basically ensures that each destination of a multicast group, that corresponds to source $i$ is scheduled at least $B_{i,j}$ slots. With these last two constraints that we explained it is possible that a multicast group is changing dynamically on a slot basis depending on the interference conditions. On the other hand constraint (5) for the power budget is also very crucial for ensuring fairness with $\mathbf{MC-ALL}$. This constraint ensures that the total power expenditure for a source during the $T$ slots is within a certain budget. Therefore, the source must comply with this power budget regardless of how many times it was activated for completing a single multicast transmission. Finally, (6) ensures that the transmit power constraint per-slot must be between a minimum and a maximum value.

The MILP is solved by relaxing the integer constraints for the slot activation indicator variables $x^{t}_{ij}$ so that a linear program (LP) is solved. The practical problem we have to address is the selection from the result of the relaxed LP, the optimal binary result that should either be 0 or 1. In works such as~\cite{fan09b} the randomized rounding method was employed for this purpose while similar approaches were followed in~\cite{tang06}. In this letter we exploit the nature of the problem, that is the multicast transmission, in order to reach the desired result. The approximation algorithm shown in Fig.~\ref{fig:scheduling-algorithm} is described next.

\begin{figure}[t]
\framebox[1.0\linewidth]{
\begin{minipage}[t]{0.9\linewidth}
$milp\_relax(\mathcal{S},\mathcal{D},T,B)$
\begin{algorithmic}[1]
\STATE $\hat{\mathbf{x}},\hat{\mathbf{P}}=lp(\mathcal{S},\mathcal{D},T,B)$ //find the solution vectors 
\FOR{all nodes $i\in \mathcal{S}$ }
\STATE $c_i=\sum^{D_i}_{j=1}\sum^{T}_{t=1} \hat{x}^{t}_{ij}$
\ENDFOR
\STATE $\mathcal{Z}=\mathcal{S}$
\FOR{all $i\in \mathcal{Z}$ }
\STATE $cnt=0$
\WHILE{$i$ not feasible}
\STATE $if(cnt==0)$ $i=\arg\max ( c_i )$, $m=\lceil c_i \rceil$
\STATE $if(cnt==1)$ $i=\arg\max ( c_i )$, $m=\lfloor c_i \rfloor$
\FOR{p2p link $j=1$ until $m$ }
\STATE $j=\arg\max \hat{x}^{t}_{ij}$
\STATE set $\tilde{x}^{t}_{ij}=1$, $\tilde{P}^{t}_{i}=\hat{P}^{t}_{i}$
\ENDFOR
\IF{DMC-OPT is feasible}
\STATE $feasible=TRUE$, $cnt=0$
\ELSE
\STATE $feasible=FALSE$, $cnt=1$
\ENDIF
\ENDWHILE ~~// Solution for $i$: $\tilde{\mathbf{x}}_{i},\tilde{\mathbf{P}}_{i}$
\STATE $\mathcal{Z}=\mathcal{Z}-\{i\}$
\STATE $\hat{\mathbf{x}},\hat{\mathbf{P}}=lp(\mathcal{Z},\mathcal{D},T,B,\tilde{\mathbf{x}}_{i},\tilde{\mathbf{P}}_{i})$
\ENDFOR ~~// The final solution: $\tilde{\mathbf{x}},\tilde{\mathbf{P}}$
\end{algorithmic}
\end{minipage}
}
\caption{Pseudocode for the approximation algorithm of the relaxed $\mathbf{DMC-OPT}$ MILP problem.}%
\label{fig:scheduling-algorithm}
\end{figure}

After the relaxed LP problem is solved (line 1) and the optimal solution is derived in the form of the vectors $\hat{\mathbf{x}},\hat{\mathbf{P}}$, the algorithm calculates the parameter $c_i=\sum^{D_i}_{j=1}\sum^{T}_{t=1} x^{t}_{ij}$ that expresses the average number of slots that source $i$ must be activated. 
The main idea of the approximation algorithm is to re-order the multicast sources according to the value of $c_i$. The source that has the highest value for $c_i$ must be scheduled for achieving throughput optimality since its transmission can reach the highest number of multicast destinations. To accomplish that, the algorithm calculates also $\lceil c_i \rceil$ which is the maximum number of links that can be active (lines 9-10). For the specific source that has the maximum $\lceil c_i \rceil$, the corresponding $x^{t}_{ij}$ for each destination are set to 1 (lines 11-14). If this schedule is feasible the algorithm selects it and moves to the next step. Otherwise it takes a number of activated point-to-point links $x^{t}_{ij}$ to be equal to $\lfloor c_i \rfloor$ (line 10) which is feasible by definition since $c_i$ corresponds to a feasible solution. The corresponding variables $x^{t}_{ij}$ are again set to 1 and are stored in the vector $\mathbf{\tilde{x}}_{i}$ while the solution for transmit power is stored in $\tilde{\mathbf{P}}_{i}$. The new LP is solved again by using as input the set $\mathcal{Z}$ of unscheduled sources and the now constant $\mathbf{\tilde{x}}_{i},\tilde{\mathbf{P}}_{i}$. 

\subsection{ILP Formulation with Constant Transmitter Power}
By considering a constant transmitter power from all the sources the problem can be significantly simplified leading to considerably interesting results even without power control. The problem is defined by removing constraint (6) and by setting $P^t_i$ to a constant power level $P$ in $\mathbf{DMC-OPT}$.
The solution to the above problem is obtained with CPLEX 12.04 in the results section.

\section{Performance Evaluation}
\label{section:performance-evaluation}
In this section we compare the performance of the proposed scheduling algorithms named $\mathbf{DMC-OPT}$ to that of the multicast scheduling algorithm $\mathbf{DMC-ALL}$ where the hypergraph of each mulitcast link is treated as one schedulable entity. We also present results for the scheduling of unicast transmissions named $\mathbf{UNI-ALL}$. The SNR threshold $\beta$ is 10dB, the maximum and minimum transmit power levels are $P^{t,max}_i$=300mW, $P^{t,min}_i$=0.01$P^{t,max}_i$, while for the ILP case $P$=0.3$P^{t,max}_i$. The node distances $d_{i,j}$ are randomly and uniformly selected is the range $[0,1]$ with the path loss exponent $a$=3. Also $\sigma^2$=0.1mW. We consider multicast groups with different number of destinations and different traffic loads in terms of the required active slots $B$.
\begin{figure}[t]
\begin{center}
\subfigure[$B$=8, $T$=8]{\includegraphics[keepaspectratio,width = 0.515\linewidth]{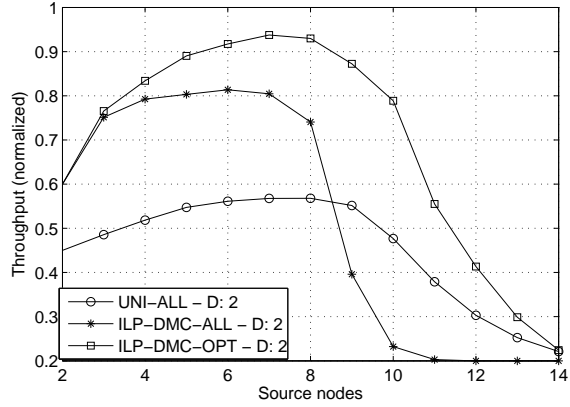}}\hspace{-0.5cm}
\subfigure[$B$=4, $T$=8]{\includegraphics[keepaspectratio,width = 0.515\linewidth]{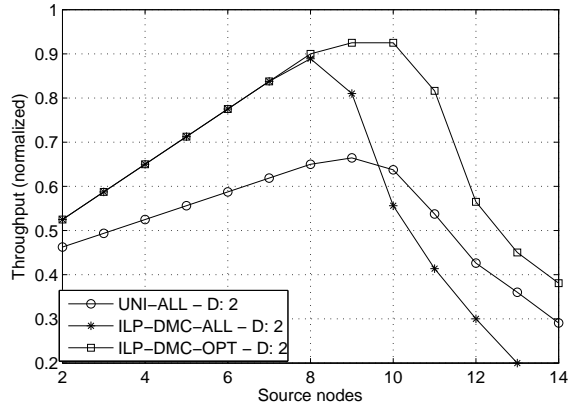}}
 \end{center}
 \caption{Results for ILP.}
 \label{fig:results1}
\end{figure}

In Fig.~\ref{fig:results1} the performance after solving the ILP is presented for both $\mathbf{DMC-ALL}$ and $\mathbf{DMC-OPT}$ with a multicast group of two destinations. In Fig.~\ref{fig:results1}(a) where $B$=$T$=8 we see that the proposed scheme leads to high throughput increase. It is important to observe that the performance of the $\mathbf{DMC-OPT}$ scheme reaches a peak at a slightly higher number of multicast sources/groups. Also note that as the number of sources is increased, the performance of $\mathbf{DMC-ALL}$ deteriorates faster than the performance of $\mathbf{UNI-ALL}$ and this is only because the increased node density increases interference. For a lighter traffic load of $B$=4 and $T$=8 in Fig.~\ref{fig:results1}(b), the performance trend is similar.

\begin{figure}[t]
\begin{center}
\subfigure[$B$=4, $T$=8]{\includegraphics[keepaspectratio,width = 0.515\linewidth]{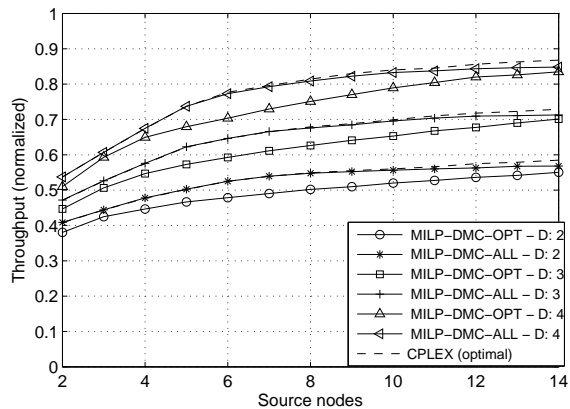}} \hspace{-0.5cm}
\subfigure[$B$=8, $T$=8]{\includegraphics[keepaspectratio,width = 0.515\linewidth]{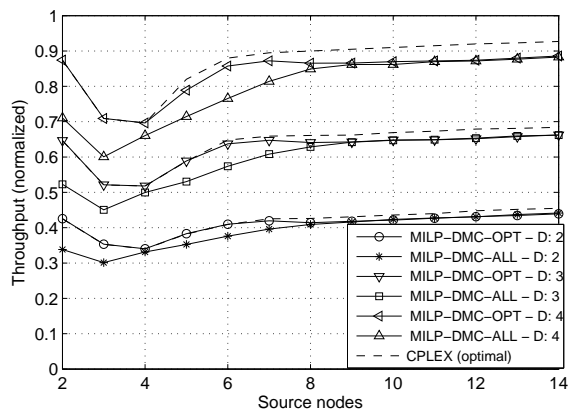}}
 \end{center}
 \caption{Results for MILP.}
 \label{fig:results2}
\end{figure}

The results for the LP relaxation of the MILP problem are shown in Fig.~\ref{fig:results2}. With dashed lines we present the optimal solution calculated with CPLEX. We observe in Fig.~\ref{fig:results2}(a) that for a traffic requirement of $B$=4 out of $T$=8 slots, there is a performance improvement for $\mathbf{DMC-OPT}$ that is increased as the number of multicast destinations is increased. This behavior occurs primarily for a number of sources that is less than the number of maximum number of used slots $T$ while for high node density interference dominates again. For a higher traffic load and backlogged nodes ($B$=$T$=8) the results can be seen in Fig.~\ref{fig:results2}(b). For different number of destinations, the performance improvement of $\mathbf{DMC-OPT}$ is higher over $\mathbf{DMC-ALL}$ even for a small number of sources. However, in both cases and for higher node density the performance of all schemes converges. The reason is that interference is higher and fewer options for scheduling exist (the activation of a source for one destination generates nearly the same interference even if more destinations are activated). From the results that represent the optimal solution obtained with CPLEX, we can see that the approximation algorithm is more sensitive to the number of multicast destinations than the actual number of sources. For a lighter traffic load the proposed algorithm can approach closer the optimal solution.

\section{Conclusions}
\label{section:conclusions}
In this letter we presented ILP \& MILP formulations for the problem of multicast link scheduling in wireless networks. With the proposed formulation a single multicast transmission is separated across different time slots while complying with the power budget. For the MILP formulation we proposed an approximation algorithm that exploits the multicast nature of the problem. The performance results indicate that as the multicast group is increased, it is more critical to employ the proposed approach that allows the scheduling algorithm to freely allocate individual transmissions across time. For constant transmit power, the proposed approach offers higher throughput benefits because with existing schemes many multicast links are disqualified from being scheduled. In our future work our plan is to derive an analytical approximation bound for the proposed algorithm with respect to the optimal solution.

\bibliography{../tony-bib}

\begin{thebibliography}{1}
\providecommand{\url}[1]{#1}
\csname url@samestyle\endcsname
\providecommand{\newblock}{\relax}
\providecommand{\bibinfo}[2]{#2}
\providecommand{\BIBentrySTDinterwordspacing}{\spaceskip=0pt\relax}
\providecommand{\BIBentryALTinterwordstretchfactor}{4}
\providecommand{\BIBentryALTinterwordspacing}{\spaceskip=\fontdimen2\font plus
\BIBentryALTinterwordstretchfactor\fontdimen3\font minus
  \fontdimen4\font\relax}
\providecommand{\BIBforeignlanguage}[2]{{%
\expandafter\ifx\csname l@#1\endcsname\relax
\typeout{** WARNING: IEEEtran.bst: No hyphenation pattern has been}%
\typeout{** loaded for the language `#1'. Using the pattern for}%
\typeout{** the default language instead.}%
\else
\language=\csname l@#1\endcsname
\fi
#2}}
\providecommand{\BIBdecl}{\relax}
\BIBdecl

\bibitem{chandra09}
R.~Chandra, S.~Karanth, T.~Moscibroda, V.~Navda, J.~Padhye, R.~Ramjee, and
  L.~Ravindranath, ``Dircast: A practical and efficient wi-fi multicast
  system,'' in \emph{ICNP}, 2009.

\bibitem{dujovne06}
D.~Dujovne and T.~Turletti, ``Multicast in 802.11 wlans: an experimental
  study,'' in \emph{MSWiM}, 2006.

\bibitem{wieselthier00}
J.~Wieselthier, G.~Nguyen, and A.~Ephremides, ``On the construction of
  energy-efficient broadcast and multicast trees in wireless networks,'' in
  \emph{IEEE INFOCOM}, 2000.

\bibitem{wang03}
K.~Wang, C.~Chiasserini, R.~R. Rao, and J.~Proakis, ``A distributed joint
  scheduling and power control algorithm for multicasting in wireless ad hoc
  networks,'' in \emph{ICC}, 2003.

\bibitem{gopinathan09}
A.~Gopinathan, Z.~Li, and C.~Williamson, ``Optimal multicast in multi-channel
  multi-radio wireless networks,'' in \emph{IEEE/ACM MASCOTS}, 2009.

\bibitem{krishnan10}
K.~R. Krishnan, D.~Shallcross, and L.~Kant, ``Joint optimization of scheduling
  and multicast trees by column-generation,'' in \emph{WiOpt}, 2010.

\bibitem{cui10}
T.~Cui, L.~Chen, and T.~Ho, ``On distributed scheduling in wireless networks
  exploiting broadcast and network coding,'' \emph{IEEE Transactions on
  Communications}, vol.~58, no.~4, pp. 1223 --1234, April 2010.

\bibitem{fan09b}
S.~Fan, L.~Zhang, and Y.~Ren, ``Approximation algorithms for link scheduling
  with physical interference model in wireless multi-hop networks,''
  \emph{CoRR}, vol. abs/0910.5215, 2009.

\bibitem{tang06}
J.~Tang, G.~Xue, C.~Chandler, and W.~Zhang, ``Link scheduling with power
  control for throughput enhancement in multihop wireless networks,''
  \emph{Vehicular Technology, IEEE Transactions on}, vol.~55, no.~3, pp. 733
  --742, May 2006.

\end{thebibliography}

\end{document}